# Interdependence in active mobility adoption: Joint modeling and motivational spill-over in walking, cycling, and bikesharing

Maher Said[a,†], Alec Biehl[b, †] and Amanda Stathopoulos[a]*

[a] Department of Civil and Environmental Engineering, Northwestern University, Evanston, IL;

[b]National Transportation Research Center, Oak Ridge National Laboratory, Knoxville, TN

[†] The authors contributed equally to the manuscript
* Corresponding author:
2145 Sheridan Road, Tech #A335, Evanston, IL 60208-3109
Phone: 847-491-5629, E-mail: a-stathopoulos@northwestern.edu

**Abstract**

Active mobility offers an array of physical, emotional, and social well-being benefits. However, with the proliferation of the sharing economy, new nonmotorized means of transport are entering the fold, complementing some existing mobility options while competing with others. The purpose of this research study is to investigate the adoption of three active travel modes —namely walking, cycling and bikesharing — in a joint modeling framework. The analysis is based on an adaptation of the stages of change framework, which originates from the health behavior sciences. Multivariate ordered probit modeling drawing on U.S. survey data provides well-needed insights into individuals' preparedness to adopt multiple active modes as a function of personal, neighborhood and psychosocial factors. The research suggests three important findings. 1) The joint model structure confirms interdependence among different active mobility choices. The strongest complementarity is found for walking and cycling adoption. 2) Each mode has a distinctive adoption path with either three or four separate stages. We discuss the implications of derived stage-thresholds and plot adoption contours for selected scenarios. 3) Psychological and neighborhood variables generate more coupling among active modes than individual and household factors. Specifically, identifying strongly with active mobility aspirations, experiences with multimodal travel, possessing better navigational skills, along with supportive local community norms are the factors that appear to drive the joint adoption decisions. This study contributes to the understanding of how decisions within the same functional domain are related and help to design policies that promote active mobility by identifying positive spillovers and joint determinants.

# 1 Introduction

The overarching goal of sustainable travel campaigns is to reduce carbon emissions arising from widespread single-occupancy vehicle use. In recognition of the strong evidence for tangible well-being and community-oriented outcomes of active mobility (Handy et al., 2014; Page and Nilsson, 2017; Singleton, 2018), research on designing behavior change interventions that encourage the adoption of these modes has flourished in the past decade. While the built environment has well-established impact on non-motorized transportation decisions, particularly via destination accessibility measures (Ewing and Cervero, 2010), recent work tends to emphasize attitudinal and social factors. For instance, the degree to which social norms influence transportation decision-making processes is the focus of many recent studies demonstrating how 'soft policy' approaches, based on behavioral economic and psychological theory, could be more effective than built environment interventions or financial incentives (Heinen and Handy, 2012; Heinen et al., 2010; Riggs, 2017).

Nonetheless, various internal (e.g. car dependence, physical fitness) and external (e.g. safety hazards, urban form) barriers could make travel by nonmotorized modes infeasible, resulting in the likely undermining of behavior change campaigns (Heinen and Ogilvie, 2016; Ruiz and Bernabé, 2014; Zuniga, 2012). These barriers, as well as motivators, of adoption are also likely to vary by mode, suggesting that analyzing 'active travel' as a general category limits the applicability and scope of derived policy conclusions. Moreover, the proliferation of micro-mobility innovation, fueled by online platforms and mobile-enabled access, is making new alternatives available to urban travelers for short-distance trip-making (Heineke et al., 2019; Zarif et al., 2019). This richer choice environment has spurred a range of questions regarding how shared travel modes might interact with equally feasible traditional options (Biehl & Stathopoulos 2020).

In accordance with the above viewpoints, this study investigates the adoption of walking, cycling, and bikesharing, with the aim to determine both specific and joint obstacles and facilitating factors. Traditionally, mode adoption has been modeled as a discrete event with covariates added to explain the status of being a user vs non-user. Here we take the view that adoption is a process by accounting for the distinct stages of adoption, a perspective rooted in the Transtheoretical Model (Prochaska & Velicer, 1997). Originally developed as a synthesis of accumulating trends in cognitive-behavioral therapy practices for smoking cessation, this model has been the subject of recent—yet limited—attention in the transportation and sustainability literatures (Forward, 2014; Gatersleben and Appleton, 2007; Langbroek et al., 2016; Waygood and Avineri, 2016). The model posits that individuals fall in one of several stages that constitute a sequence of readiness for behavior change, and that successful progression through the stages towards a desirable behavioral outcome—such as sustainable travel mode adoption—requires the implementation of stage-specific processes of change. For example, encouraging individuals to consider a new behavior would require effective dissemination of information regarding its benefits, while habit formation necessitates *helping relationships* to support the desired change, the source of which could be an individual's social network (Friman et al., 2017).

It is likely that active travel choices function as either complements or substitutes to one another, so ignoring these relationships in simple regression-based modeling could lead to inaccurate representations of potential modality style changes. Therefore, it is important to capture the correlation between, for instance, the level of adoption of walking in tandem with cycling, to study behavior change readiness and determine whether these decisions are jointly or independently motivated. Moreover, this allows a deeper study of common determinants of adopting different active modes that would yield important insights for practical promotion efforts. The contributions of this paper are therefore threefold. First, it expands on the exploration of *jointness* of adoption decisions in the travel behavior literature via multivariate ordered probit modeling where three active modes -- walking, biking and bikesharing -- are studied as dependent variables. Second, it expands the growing body of socio-psychological inquiries into active travel adoption through stage-based frameworks. Specifically, the active mobility adoption is modelled as a series of stages where membership is treated as an ordinal outcome. Third, we identify joint determinants that underpin adoption-decisions to elucidate the complementarity among mode decisions. Connecting the stage-of-change analysis with joint modeling of active modes affords distinct insights into the process of change, and how the adoption process varies across similar modes. The multivariate model results reveal both *connections*, namely joint determinants, and *distinctions*, namely that each mode has a distinct adoption path, that lead to new practice insights.

## 2 Literature Review

### *2.1 Multi-Stage Behavior Change Theory for Active Travel*
A growing body of knowledge has theorized and tested stages-of-change analysis to explain adoption of active travel. From a broad conceptual standpoint, Scheiner (2018) details how the social context and networks, habitual behavior, lifestyle, and major life events are key pillars of the behavior change process. Anagnostopoulou et al. (2018) overview existing persuasive strategies for inducing change in mobility patterns, such as social comparison to foster competitiveness among participants. Meanwhile, Andersson et al. (2018) propose a framework for designing behavior change support systems that combines the following four theories: Diffusion of Innovations, Transtheoretical Model (TTM), Gamification, and Theory of Planned Behavior. This latter review emphasizes that travel behavior change campaigns should be theory-driven to allow for comparison of findings across contexts. That being said, theoretical integration and the process of choosing the most appropriate theory to guide policy and practice are still greatly debated among researchers (Bamberg, 2013; Hagger and Chatzisarantis, 2009). Nonetheless, one of the primary reasons why the Transtheoretical Model (Prochaska and Velicer, 1997), and the more recent Stage-Model of Self-Regulated Behavioral Change (Bamberg, 2013), are garnering recent attention as effective tools for promoting pro-environmental behavior such as car use reduction and sustainable mobility adoption (Handy et al., 2014; Keller et al., 2019) is their focus on understanding the process of behavioral change. Additionally, multi-stage frameworks have a greater capacity to capture the translation of intention into behavior compared to the more widely-used Theory of Planned Behavior (Ajzen, 1991) and Value-Belief-Norm Theory (Stern et al., 1999), which have been criticized for their lack of explanatory power on this matter. Accordingly, the corresponding statistical models that delineate stage-specific determinants of readiness for—or resistance to—change offer more reliable guidance on the design of policy interventions (Winters et al., 2011).

It is imperative, however, that transportation researchers be aware of conceptual and empirical shortcomings within the source literature. Importantly, in a comprehensive review of the Transtheoretical Model, Armitage (2009) concludes the following: (a) there is a lack of longitudinal research to demonstrate the benefits of employing a multi-stage framework over a two-phase model that only considers whether an individual does or does not perform a specific behavior—similar to the discrete choice experiment approach for investigating traveler decision-making—and (b) the proposed cognitive-behavioral *processes of change* in the original framework are noticeably understudied as design principles for stage-specific interventions. On the latter point, only a few transportation research studies explicitly address this component of the TTM (Biehl et al., 2018, 2019; Parkes et al., 2016), though research by Thigpen et al. (2015) finds that attitudinal variables outperform barriers, travel attributes, and personal characteristics in statistical models of campus cycling adoption. Example processes include *consciousness raising*, defined as the extent to which people integrate information about a new behavior, and *social liberation*, defined as the realization that the social norms are evolving in support of behavioral change (Prochaska et al., 2008). Thus, despite the noted shortcomings, the TTM advances a novel market segmentation approach that, with proper data collection and analysis, could be implemented and investigated to more precisely determine individuals' orientations towards changing their travel behaviors and establish strategies that are better aligned with their motivations, opportunities, and abilities to change.

Alternative theoretical lenses have been proposed in more recent literature on active travel behavior, illustrating not only the myriad of potential factors influencing travel-related decisions but also the challenge of selecting the most appropriate framework to meet research objectives. Schneider (2013) proposes the Theory of Routine Mode Choice Decisions for walking and bicycling, which conjectures that the most effective strategy to promote these modes is the following five-step sequence of increasingly persuasive states: (1) awareness and availability, (2) basic safety and security, (3) convenience and cost, (4) enjoyment, and (5) habit. Meanwhile, Willis et al. (2015) and Cass and Faulconbridge (2016), through a rigorous examination of cycling for commuting trips, illustrate that the physical built environment factors only 'scratch the surface' when it comes to understanding mobility patterns. In particular, social practices within the household, workplace (Willis et al., 2015) and society in general (Cass and Faulconbridge, 2016) give rise to the perceived competences and meanings associated with different modes of travel, such as notions of freedom and privacy attached to car use and environmental and health attitudes related to cycling.

### *2.2 Modeling of Joint Outcomes in Transportation Research*

Transportation research has long recognized the need for joint model frameworks to account for interconnected choice dimensions, such as the fundamental relationship between residential location and mode choice or vehicle fleet size (Bhat and Guo, 2007; Pinjari et al., 2011). Studying multiple dependent variables simultaneously has important advantages to single model frameworks: it explicitly accounts for the interdependency of choices, reveals unobserved factors that affect choice dimensions simultaneously, and allows researchers to accurately parse the influence of exogenous variables (Pinjari et al., 2008; Rashidi et al., 2011; Salon, 2015). Notably, a comprehensive framework for joint estimation with mixed dependent variables (including ordered) under different functional forms is proposed by Bhat (2015). While other frameworks, most notably structural equation modeling, can also be utilized to model the interconnectedness of multiple decisions (Astroza et al., 2017; Kamargianni and Polydoropoulou, 2013; Singleton, 2018), we find that a multivariate ordered probit model best captures the essence of a stages of change analysis and is capable of pinpointing the joint determinants of adoption and motivational spillover. Thus, the remainder of this section is dedicated to reviewing recent applications in the transportation literature to demonstrate the value and insights derived from capturing 'jointness' among related travel behaviors. Characterizing and quantifying this type of relationship is becoming increasingly important with the emerging Mobility-as-a-Service phenomenon (Mulley, 2017), which depends on understanding the complementarity or substitutability of alternatives comprising a mobility service package.

Beginning with *bivariate* ordered models, Yamamoto and Shankar (2004) develop a joint probit model to examine the joint driver-passenger injury severity for single-vehicle crashes. The joint model reveals a significant correlation between the dependent variables as well as increased parameter efficiency; univariate modeling would have overlooked the joint unobserved factors connecting drivers and passengers. Anastasopoulos et al. (2012) utilize a bivariate ordered probit model to investigate joint household automobile and motorcycle ownership in Athens, Greece. The results point to a significant *negative* cross-equation correlation of car-motorcycle error terms, supporting the notion of mode substitutability in multi-vehicle households. Additionally, many household and residential context variables had varied effect for car *vs*. car-motorcycle ownership. Guo et al. (2007), meanwhile, investigate the relationship between motorized and nonmotorized mode use as an outcome of improvements to the built environment. Their model indicates that "increased bikeway density and street network connectivity have the potential to promote more nonmotorized travel to supplement individuals' existing motorized trips" (p. 1).

Other studies utilize *multivariate* probit models to explore the relationship of multiple seemingly related travel behaviors. Recognizing that individuals frequently consider multiple travel strategies simultaneously, Choo and Mokhtarian (2008), investigate several travel-related strategies, namely (a) maintenance and increase of travel, (b) reduction in travel, and (c) major location or lifestyle changes in a joint framework. The authors found the observed correlations among the three strategies to be significant, therefore justifying the validity of the multivariate framework. Golob and Regan (2002) use California-based survey data to construct a multivariate probit model that identifies attributes impacting the propensity of the trucking industry to adopt a handful of available information technologies. Joint estimation among seven presented technologies within the choice model points to simultaneous adoption of pairs of technologies by companies, again demonstrating how services might be packaged together for additional user benefit. Dias et al. (2020) study in person versus online engagement with six different shopping activity types in a multivariate ordered probit. The authors reveal bundling of the decisions as well as evidence for both complementarity and substitution effects depending on covariates. Nazari et al. (2018) estimate a multivariate ordered probit model on the level of interest in private autonomous vehicles alongside four different shared autonomous vehicle options, revealing strong positive correlation. Becker et al. (2017) examine car ownership, car-share enrollment, and two types of transit subscriptions—namely *restricted* and *full* access to the network—to capture shared attributes underlying decision-making processes. Results suggest that 'mobility portfolios' should be constructed based on travelers' attitudinal dispositions as well as situational variables that either permit flexibility or induce restrictions on feasibility sets for mode choice. Specifically, car-share enrollment appears to strongly complement restricted transit access. Finally, Tang et al. (2018) show that, during high-speed rail trips, various activities jointly contribute to the 'positive utility' of travel, depending on the trip purpose.

All in all, the cited research shows that joint estimation of naturally interdependent decisions improves model quality compared to treatment of outcomes as univariate independent phenomena, revealing otherwise overlooked insights for crafting tailored policy measures that foster changes in travel behavior with a shared target outcome (e.g. improved health).

## 3 Survey and Mobility Pattern Analysis

### 3.1 Survey Method

The survey was designed using Qualtrics software and participants were recruited from the Amazon MTurk platform. Given the focus on active mobility behavior, the sampling is restricted to major metro areas in six Midwestern states—Indiana, Illinois, Michigan, Minnesota, Ohio, and Wisconsin—as quasi-control for climate and lifestyle values. Two pilots were carried out in December 2016 to assess the survey tool, attitude constructs and completion time. Final implementation took place in February 2017. The timing was chosen to allow six months of mode-change horizon to coincide with more general active mobility viability in spring and summer. 1,253 responses were collected, and after excluding respondents in rural areas, suspicious response patterns, incomplete responses and failed attention check, 914 (73%) responses were retained. The median completion time was 14 minutes. More details on the survey instrument are provided in Biehl et al. (2018). The importance of the multimodality indicator in this study led to further reduction of the sample size to 826 respondents (66%).

The survey collected socio-demographic, travel behavior, psycho-social, and geographic information to assess which variables are valuable for distinguishing among levels of readiness to adopt three active travel modes: walking, cycling, and bikesharing. Table 1 summarizes important descriptive statistics. It is important to note that several responses were made optional to preserve privacy/sensitivity. This caused some instances of nonresponse rates. The lack of zip code information in 6.8% leads to a challenge in extracting further built environment information from an auxiliary resource (e.g. Google Places API). That being said, since the measurement of active travel behavior is not restricted to home-based trips and the fact that zip codes may exhibit considerable geospatial heterogeneity due to their size, the various 'subjective' built environment variables captured in the survey are sufficient for the intended analyses (Ma et al., 2014).

**HERE** *Table 1. Key characteristics of the survey sample (n = 826).*

Additionally, while research on the validity and quality of online sampling is still ongoing, current evidence indicates that online samples are biased towards male, younger, more educated, and wealthier respondents (Kwak and Radler, 2002; Lindhjem and Navrud, 2011). Nonetheless, all survey modes have challenges and there seems to be sufficient evidence that MTurk respondents are demographically more diverse than traditional samples (Mason and Suri, 2012; Smith et al., 2015) and that MTurk data is comparable in quality to student or other online panels (Garrow et al., 2020; Kees et al., 2017). More importantly, there is consistency in behavioural responses compared to established survey modes (Lindhjem and Navrud, 2011; Sheehan, 2018). Taken together with the piloting, screening tests and quality checks, along with timing and state inclusion criteria, we have confidence in the MTurk sample to investigate the proposed active mobility adoption behavior.

### 3.2 Stage of Adoption Analysis

Each survey respondent is assigned to a stage of change for each travel mode investigated based on indirect assignment questions (see Table A1 in Appendix). Originally, the employed staging algorithms followed the Transtheoretical Model and sorted individuals into one of six stages, but due to some small membership totals, we merged adjacent stages to produce four-stage frameworks for walking, cycling, and bikesharing adoption. The resulting stages are: Precontemplation (PC), Contemplation (C), Preparation (P), and Action-Maintenance (A-M). The identification steps between walking/cycling and bikesharing differ, however, since the latter is an emerging form of shared mobility that must be supplied by a service provider, as opposed to being more fully within an individual's control to use. The stage membership breakdown is displayed in Figure 1. A key observation is that Precontemplation has the highest membership while Preparation has the lowest membership in all frameworks. Another point worth noting is that, in the case of bikesharing, it is possible that respondents belonging to either of the two middle stages may not have access to this mode, which might seem unrealistic from a practical behavior standpoint. Although this would be a valid concern, the stage-of-change assessment was purposefully designed to demonstrate that motivation to engage in a specific behavior does not necessarily align with the existence of opportunities to do so. Therefore, from a broader perspective, defining stages in this manner would be useful for identifying the most likely 'early adopters' of a new bike share system, since it is to be expected that exposure to information and the possession of certain physical skills and psychological traits, among other things, would still influence the degree of contemplation or preparedness.

**ABOUT HERE** *Figure 1. Stage-of-change membership breakdown for three active travel modes (n = 826).*

### 3.3 Multimodal Travel Diary Analysis

To collect information regarding travel habits, respondents completed a 'weekly travel diary' representing the number of trips (five levels of frequency) typically made across eight travel modes and three trip purposes, for the mode that covered the greatest distance of any routine trip. The trip-making breakdown is described in detail in Biehl and Stathopoulos (2020). We note that that drive alone is the most frequently used mode across all trip purposes, in contrast to bikesharing which is used infrequently. Cycling use is concentrated in the leisure category, while walking is the most generally used mode outside of driving, used for both shopping and leisure trips. The multimodal travel diary reveals different mode patterns than the above adoption stage classification. The difference is due to the narrower definition and time-frame used in the travel diary, focusing only on the mode used for the greatest distance of a trip, and a weekly time-span.

The trip diary is designed to measure the degree of multimodality. For modeling purposes, first the weekly trip frequency categories are converted into continuous values using the midpoint of the frequency band. Next the modified Shannon entropy index (SEI) is computed to represent variability in the observed mode frequency based on recommendations by Diana and Pirra (2016). Equation 1 shows the SEI, where $n$ is the total number of modes considered, $M$ is the maximum reported frequency across all modes and $f_i$ is the frequency of the $i$th mode. The index falls along the unit interval, where values closer to one indicate stronger multimodality.

$$\text{SEI} = \sum_{i=1}^{n} \left\{ \frac{f_i}{nM} \left[ 1 + \ln\left(\frac{M}{f_i}\right) \right] \right\} \tag{1}$$

Another indicator of multimodality, Herfindahl-Hirschman Index, was evaluated. Nonetheless, due to the conceptual similarity, only the modified Shannon entropy index was included in the final statistical models to avoid multicollinearity.

### 3.4 Psychological Factors

Exploratory factor analysis (EFA) was applied to explore the general constructs of *Active Travel Disposition* and *Environmental Spatial Ability*, namely the psychosocial constructs most likely to affect active mode decisions. The factor identification builds on work featured in Biehl et al. (2019), whereas the current EFA solutions advanced on the previous work in three ways. First, we removed items whose loadings were below 0.45 after varimax rotations, resulting in three new single-item measures in addition to the ones comprising travel satisfaction and built environment perceptions scales. Second, the number of factors per individual scale remains the same apart from *Active Travel Disposition*, which now consists of two separate identity constructs as well as one representing personal norms; that is, two (overlapping) classes of identity-related items, namely how active travel interventions might impact (a) the individual and (b) the local community, are distinguished. Third, factor scores were calculated using all scale items, rather than just those with loadings meeting the specified threshold requirement, employing the tenBerge scoring method as explained in R's *psych* package documentation (Revelle, 2016). For the sake of brevity, we save the discussion of factor interpretations for Section 5, restricted to only those statements and constructs appearing as significant in the final models. The items comprising each factor are included in Tables A2 to A5 in the Appendix, including full formulations of the psychometric statements defining each item.

## 4 Statistical Methods

Multivariate ordered regression models are a natural extension of their univariate counterparts, where the coefficients for different behaviors are simultaneously estimated with an unrestricted correlation matrix of the random error terms (Greene and Hensher, 2010). In this study, the multivariate specifications are used to capture the potential connections between the stages of adoption for travel by privately owned bike, shared bike and walking.

The ordinality of the dependent variable is accounted for by a latent continuous variable $y_i^*$ which is defined through a censoring approach as follows:

$$y_i = \begin{cases} y_1 & if & y_i^* \leq \mu_1 \\ y_2 & if & \mu_1 < y_i^* \leq \mu_2 \\ \vdots & \vdots & \vdots \\ y_{J-1} & if & \mu_{J-1} < y_i^* \leq \mu_J \\ y_J & if & \mu_J < y_i^* \end{cases} \tag{2}$$

Here, $y_i^*$ is a continuous latent variable, $y_i$ is the ranked (or ordinal) choice observed, $J$ is the number of ordered ranks (or stages), and $\mu_1$ to $\mu_J$ are a vector of threshold parameters to be estimated (Eq. 2). It is also assumed that the latent variable $y_i^*$ is unbounded, with $\mu_0$ being $-\infty$ and $\mu_{J+1}$ being $+\infty$. The resulting latent regression model has the familiar structure (Eq. 3),

$$y_i^* = \boldsymbol{\beta}' \boldsymbol{x}_i + \varepsilon_i, \text{ where } i = 1, \dots, n \tag{3}$$

where $n$ is the total number of individuals, $\boldsymbol{\beta}$ is a vector of coefficients, $\boldsymbol{x_i}$ are independent variables and $\varepsilon_i$ is the error term of a specified distribution, typically either logistic or normal. The resulting probability on which the log-likelihood is estimated is presented in equation 4:

$$P(y_i = j | \boldsymbol{x}_i) = F(\mu_j - \boldsymbol{\beta}' \boldsymbol{x}_i) - F(\mu_{j-1} - \boldsymbol{\beta}' \boldsymbol{x}_i), \text{ where } j = 0, 1, \dots, J \tag{4}$$

Here, $F$ is the selected cumulative distribution function (CDF), typically either logistic for an ordered logit or normal for an ordered probit model.

It is important to restrict every $\mu_{j-1}$ to be less than $\mu_j$ to ensure that the above probability is positive for every $j$. Extending to trivariate ordered regression models using the probit form, the multiple equation specification is shown in equation 5:

$$\begin{aligned} y_{i,1}^* &= \boldsymbol{\beta}_1' \boldsymbol{x}_{i,1} + \varepsilon_{i,1}, & \text{where} & & y_{i,1} = j & \text{if } \mu_{j-1,1} < y_{i,1}^* < \mu_{j,1} & \text{and where } j = 0, \dots, J \\ y_{i,2}^* &= \boldsymbol{\beta}_2' \boldsymbol{x}_{i,2} + \varepsilon_{i,2}, & \text{where} & & y_{i,2} = k & \text{if } \mu_{k-1,2} < y_{i,2}^* < \mu_{k,2} & \text{and where } k = 0, \dots, K \\ y_{i,3}^* &= \boldsymbol{\beta}_3' \boldsymbol{x}_{i,3} + \varepsilon_{i,3}, & \text{where} & & y_{i,3} = l & \text{if } \mu_{l-1,3} < y_{i,3}^* < \mu_{l,3} & \text{and where } l = 0, \dots, L \\ & & & (\varepsilon_{i,1}, \varepsilon_{i,2}, \varepsilon_{i,3}) \sim N(\boldsymbol{0}, \boldsymbol{R}) \end{aligned} \tag{5}$$

where $\boldsymbol{R}$ is an unrestricted correlation matrix of the random terms, and $K$ and $L$, similarly to $J$, are the number of ranks for their respective dependent variables (Greene and Hensher, 2010). In a trivariate setting, the correlation matrix $\boldsymbol{R}$ is a 3x3 matrix as presented in equation 6.

$$\boldsymbol{R} = \begin{pmatrix} 1 & \rho_{12} & \rho_{13} \\ \rho_{12} & 1 & \rho_{23} \\ \rho_{13} & \rho_{23} & 1 \end{pmatrix} \tag{6}$$

The resulting joint probability, which is the probability that enters the log likelihood for estimation, is developed in equation 7:

$$
\begin{aligned}
P\big(y_{i,1}, & y_{i,2}, y_{i,3} \big| \boldsymbol{x}_{i,1}, \boldsymbol{x}_{i,2}, \boldsymbol{x}_{i,3}\big) = \\
& \Phi_3\big[\big(\mu_{j+1,1} - \boldsymbol{\beta}_1'\boldsymbol{x}_{i,1}\big), \big(\mu_{k+1,2} - \boldsymbol{\beta}_2'\boldsymbol{x}_{i,2}\big), \big(\mu_{l+1,3} - \boldsymbol{\beta}_3'\boldsymbol{x}_{i,3}\big), \rho_{12}, \rho_{13}, \rho_{23}\big] \\
& - \Phi_3\big[\big(\mu_{j,1} - \boldsymbol{\beta}_1'\boldsymbol{x}_{i,1}\big), \big(\mu_{k+1,2} - \boldsymbol{\beta}_2'\boldsymbol{x}_{i,2}\big), \big(\mu_{l+1,3} - \boldsymbol{\beta}_3'\boldsymbol{x}_{i,3}\big), \rho_{12}, \rho_{13}, \rho_{23}\big] \\
& - \Phi_3\big[\big(\mu_{j+1,1} - \boldsymbol{\beta}_1'\boldsymbol{x}_{i,1}\big), \big(\mu_{k,2} - \boldsymbol{\beta}_2'\boldsymbol{x}_{i,2}\big), \big(\mu_{l+1,3} - \boldsymbol{\beta}_3'\boldsymbol{x}_{i,3}\big), \rho_{12}, \rho_{13}, \rho_{23}\big] \\
& - \Phi_3\big[\big(\mu_{j+1,1} - \boldsymbol{\beta}_1'\boldsymbol{x}_{i,1}\big), \big(\mu_{k+1,2} - \boldsymbol{\beta}_2'\boldsymbol{x}_{i,2}\big), \big(\mu_{l,3} - \boldsymbol{\beta}_3'\boldsymbol{x}_{i,3}\big), \rho_{12}, \rho_{13}, \rho_{23}\big] \\
& - \Phi_3\big[\big(\mu_{j,1} - \boldsymbol{\beta}_1'\boldsymbol{x}_{i,1}\big), \big(\mu_{k,2} - \boldsymbol{\beta}_2'\boldsymbol{x}_{i,2}\big), \big(\mu_{l,3} - \boldsymbol{\beta}_3'\boldsymbol{x}_{i,3}\big), \rho_{12}, \rho_{13}, \rho_{23}\big] \\
& + \Phi_3\big[\big(\mu_{j,1} - \boldsymbol{\beta}_1'\boldsymbol{x}_{i,1}\big), \big(\mu_{k,2} - \boldsymbol{\beta}_2'\boldsymbol{x}_{i,2}\big), \big(\mu_{l+1,3} - \boldsymbol{\beta}_3'\boldsymbol{x}_{i,3}\big), \rho_{12}, \rho_{13}, \rho_{23}\big] \\
& + \Phi_3\big[\big(\mu_{j,1} - \boldsymbol{\beta}_1'\boldsymbol{x}_{i,1}\big), \big(\mu_{k+1,2} - \boldsymbol{\beta}_2'\boldsymbol{x}_{i,2}\big), \big(\mu_{l,3} - \boldsymbol{\beta}_3'\boldsymbol{x}_{i,3}\big), \rho_{12}, \rho_{13}, \rho_{23}\big] \\
& + \Phi_3\big[\big(\mu_{j+1,1} - \boldsymbol{\beta}_1'\boldsymbol{x}_{i,1}\big), \big(\mu_{k,2} - \boldsymbol{\beta}_2'\boldsymbol{x}_{i,2}\big), \big(\mu_{l,3} - \boldsymbol{\beta}_3'\boldsymbol{x}_{i,3}\big), \rho_{12}, \rho_{13}, \rho_{23}\big]
\end{aligned}
\tag{7}
$$

where $\Phi_3$ represents the joint cumulative distribution function (Scott and Kanaroglou, 2002). In addition to the above references, we refer readers to Washington et al. (2020) and Bhat and Srinivasan (2005) for a more detailed treatment of ordinal models.

## 5 Modeling Results

The main goal of this study is to uncover the potential connections among the stages of decision-making to travel by various active modes, namely walking, privately-owned bike, and shared bike. These mode adoption dynamics are captured here through the correlation among the decision-making processes comprising the active mode triad. Model estimation utilizes R's *mvord* package (Hirk et al., 2020), which calculates full-information maximum likelihood estimates of joint multivariate regression models.

As a reference, separate univariate models are estimated using Stata's *oprobit* function (StataCorp, 2017) and refined independently. This ensures fundamental understanding of which variables are significant for each of the adoption processes, prior to the inclusion of correlation terms representing joint dynamics among alternatives. The variable nomenclature and definitions are provided in Table 2.

***ABOUT HERE*** *Table 2. Names and definitions of probit model covariates besides the extracted factor variables.*

### 5.1 Univariate Ordered Base Models: separate mode adoption analysis

Table 3 presents the single mode ordered probit models used as reference to determine significant covariates of each stage-of-change process. These models result from extensive specification testing and includes only significant explanatory coefficients along with the complete set of stage thresholds $\mu_j$. These reference models reflect the initial assumption that active mode decisions are independent processes. Before turning to examine the joint model specification in Sec. 5.2., we observe some general patterns from Table 3.

***ABOUT HERE*** *Table 3. Independently Estimated Ordered probit models for biking, walking and bike-share.*

Generally, the cycling and walking models have better fit and are explained by a larger number of covariates compared to bikesharing. The results also suggest stronger parallels between walking and cycling, with a balanced impact of both tangible sociodemographic and psychological factors. Instead, bikesharing is predominantly connected to psychological and neighborhood variables and largely unaffected by sociodemographic ones. The decisions to travel more by private bicycle and walking are affected by the number of vehicles in a household (compared to not owning any private cars). These observations are in line with findings by Cervero and Duncan (2003). The number of bicycles in a household, as expected, favors cycling, while not having any significant impact on the choice to use bikesharing. The lack of a driver's license results in a higher propensity for walking, as evidenced in the literature (Clark and Scott, 2013; Copperman and Bhat, 2007). Living in suburban, compared to urban or hybrid areas, decreases the likelihood of making trips by private bicycle or walking (Pucher and Dijkstra, 2000; Pucher et al., 2011). Men tend to cycle more than women, while no gender effect is found for higher frequency use of shared bicycle or walking (Heinen et al 2010). Employment and education, on the other hand, influence the decision-making stages for walking trips, with the coefficient for full-time employment being negative whereas that of higher education is positive. Unsurprisingly, more flexibility in work-schedules favors utilization of bikesharing. As for the adjusted Shannon entropy index, this is positive and significant in all three models, indicating that each active travel mode is associated with multimodal behavior, though most strongly for both cycling cases. The impact of psychological and neighborhood variables will be analyzed in more depth in relation to the multivariate model. We specifically highlight the role of identity and norms in shaping mode complementarity.

Taking a closer look at the threshold values, we observed that for each mode one of the threshold estimates is insignificant at conventional levels of significance. Specifically, this implies that cycling, be it private or shared, should be defined as a 3-stage adoption process. For walking there is stronger evidence for a distinct 4-stage adoption process.

Two modeling caveats related to variable definitions need to be discussed. First, when assigning users to their bike-sharing stage, we do not explicitly account for the local availability of a bike-share system (see Sec 3.2). By relying on the user stated intentions we give up some realism to better align modeling with the motivational focus inherent in early consideration of change. Second, the psychosocial variables were formulated in relation to general active mobility, not specific modes. This broader approach reveals a general predisposition towards active travel and is more parsimonious than mode-specific attitude constructs. We expect that our broader take on these specification issues likely leads to 'omitting' some explanatory power in our modeling.

### 5.2 Multivariate Ordered Model: joint mode adoption analysis

Table 4 shows the final trivariate model structure that jointly examines adoption of cycling, bikesharing, and walking. This structure advances on the earlier work in two main ways: 1) it accounts for potential correlation among the mode decisions, 2) it accounts for mode-specific stages-of-change thresholds.

***ABOUT HERE*** *Table 4. Trivariate probit model for stages-of-change for all three modes.*

The multivariate specification where active mode adoption is treated jointly, broadly confirms the findings from the separate structures. Yet, the trivariate structure outperforms the separate modelling in several ways: in terms of model fit, by providing more robust findings to study joint determinants, and most importantly, by generating new insights on the correlation among active travel adoption processes. Before analyzing mode-adoption relatedness (sec. 5.2.1-5.2.3), joint determinants (Sec. 5.2.4) and threshold applications (Sec. 5.2.5) in more depth, some general observations are called for. Regarding model fit, with an $\rho^2$ of 0.149, the trivariate model provides a significantly better fit than any of the three univariate models according to the Horowitz' $\rho^2$ test. While provided for completeness, the AIC and BIC are not used to compare across the univariate and multivariate models as these fit indices cannot be used to compare structures with distinct outcome variables. In terms of explanatory parameters, the findings closely match the univariate models with a few exceptions. The private biking model component closely mirrors the separate reference model with the same set of variables, identical signs and closely matching coefficient magnitudes (on average just 4% lower). Bikesharing remains related to only six explanatory variables in the multivariate structure, with no sign inversions but larger shifts in coefficient magnitude (on average 10% lower). Finally, walking is confirmed as the most articulated decision structure with closely matching coefficients (averaging a 2% difference) except for the multimodality index which is no longer significant.

The model results are discussed below in terms of mode pairs, emphasizing the degree of correlation and complementarity in causal factors. Section 5.2.5. analyzes the threshold estimates in more detail, and provides an application of the thresholds as contours of change.

*5.2.1 The active travel foundation: Cycling vs. Walking*

The joint model clearly indicates a complementarity between these two modes. The error term correlations are also the strongest of the three correlation coefficients ($\rho = 0.439$), likely reflecting the more immediate control respondents have over cycling and walking adoption compared to the uptake of a mobility service platform such as bike-share. The foundational ties between cycling and walking are related to both tangible factors (urban density, household vehicle ownership) and psychological traits (self identity). Specifically, the active mobility identity seems to promote this mode-dyad by fostering a social identity where diffusion of active travel in the local community reflects back on respondent's aspirations (Bamber et al. 2007, Heinen & Handy 2012).

Some distinctions between walking and cycling adoption also need to be made. The perceived quality of the built environment only affects walking propensity, along with having a good mental map of the neighborhood. These features are all tied to spatial navigation (Biehl et al. (2019). Instead, biking is related to multimodality, openness to learning and variety seeking, suggesting it is more connected to being open to change and learning new skills. Yet, there is a unifying theme to note. Both the observed confidence in having reliable mental maps among pedestrians, and the greater openness to learning new behaviors of cyclists relate to the *consciousness raising* process of change (Prochaska et al., 2008). This suggests that travel behavior change campaigns should focus on promoting both forms of active mobility jointly, where cycling would cover longer trip distances within specified time constraints compared to walking.

### 5.2.2 Competing biking options? Cycling vs. Bikesharing

Understanding how different biking opportunities affect active mobility adoption centers on parsing whether substitution or complementarity is at play. The strong similarity of the two biking modes would lead us to suspect a strong negative correlation, justified by filling the same mobility needs for people. On the contrary, a modest positive error correlation is revealed ($\rho = 0.181$). This positive correlation suggests a complementary relationship between private and public bike usage. This is a valuable result suggesting that ownership and habitual use of a private bike does not diminish interest in shared bike adoption. This complementarity between different forms of biking is also observed by Castillo-Manzano et al. (2015) and Biehl et al. (2019). Further analysis of longitudinal observations would be needed to reveal if this 'jointness' between biking adoption decisions is triggered by private biking leading to more openness to use public bikes, or vice-versa. Interestingly, the joint biking decisions are driven more by psychological factors than by socio-demographics of users compared to the other dependent variables in the model. The most important joint determinants relate to multi-modality, an active mobility identity grounded in both self-concept and the local community, and an orientation towards travel variety. Particularly, in line with longitudinal observations by Heinen (2018), more multimodal mobility styles help individuals move up the adoption ladder for both, private and shared biking. The results suggest that agencies and communities seeking to promote cycling will need to heed the symbolic nature of the decision to use both owned and shared bicycles, and the local community contexts. The path towards more established biking habits, either as independent or joint processes, appear to be driven by similar factors centered on identity and adaptable mobility styles.

### 5.2.3 Separate tracks: Bikesharing vs. Walking

When examining the separate models, shared bike use and walking appear to be the least overlapping processes. These two modes exhibit a very modest, yet significant, *negative* correlation of value -0.097. The finding is intuitive given that these two modes are the furthest apart in terms of skills and ownership structure, and the significant error correlation suggests slight substitutability of these modes for leisure trips. Hence, the critical cognitive and behavioral processes associated with adoption of bikesharing and walking operate almost as independent choices.

### 5.2.4 Connective tissue: Identity and norms

An important finding that emerges from the joint model is that identity and norms act as a buttress for active mobility adoption, with more consistent links than what is found for the socio-demographic variables. We observe that *self-identity* (i.e. the concept of seeing active mobility as a reflection of oneself and embodied ideals) is significant for all three modes, implying that utilizing active travel modes is in part a consequence of identity-behavior congruence that is evidenced in previous studies related to specific (Fielding et al., 2008; Johe and Bhullar, 2016) and general (Carfora et al., 2017; van der Werff et al., 2013; Walton and Jones, 2018; Whitmarsh and O'Neill, 2010) pro-environmental behaviors. In addition, *place identity*, which according to Uzzell et al. (2002) "describes a person's socialization with the physical world" (p. 29) relates to identification, cohesion and satisfaction with place, appears in both biking models. This finding suggests that individual-environment congruence, an important foundation for well-being (Knez and Eliasson, 2017; Moser, 2009), is fundamental to the adoption of habitual cycling behaviors, which may be due to the fact that this mode is not traditionally associated with dedicated 'travel space' in comparison to walking and driving. Meanwhile, *social identity*, which Uzzell et al. (2002) describe as "emotional meanings attached to identification with a social group" (p. 29), is significant only for walking. One plausible explanation for this finding is that individuals value routine pedestrian behavior because they experience stronger cohesion with local community members (French et al., 2014; Wilkerson et al., 2012), which translates into stronger identification with neighborhood-based activities. Finally, *personal norm* development is also critical for walking and bikesharing, thus aligning with other research studies demonstrating the importance of moral-based decision-making in relation to sustainable travel behaviors and commitment to making changes towards achieving a higher purpose (Bamberg et al., 2007; Chorus, 2015; Keizer et al., 2019; Klöckner and Blöbaum, 2010; Lind et al., 2015).

### 5.2.5 Threshold application and stage-of-change contours

The use of an ordered probability model is justified by the view that active mobility adoption occurs in ordered stages. Yet, beyond the estimation of the µ coefficients to represent stage-thresholds, the interpretation remains elusive (Washington et al. 2020). To provide more practical insights, this section illustrates the meaning and implied response of those thresholds in several probability contour prediction scenarios.

*ABOUT HERE* Figure 2. Threshold structure for active mode adoption processes (Multivariate structure)

Originally, the dependent variables (stages-of-change) were all assumed to have 4 stages as identified from the survey tool. However, in the course of modeling, some thresholds were found to be insignificant, resulting in both cycling and bikesharing being condensed to 3-stage processes (Figure 2). Specifically, *contemplation* and *preparation,* namely the intermediate adoption stages namely, were merged for cycling, and *precontemplation* and *contemplation*, the earliest adoption stages, for bikesharing. As a result, cycling is delineated by the following stages, (1) pre-preparation [merged], (2) preparation, and (3) action-maintenance. The stages for bikesharing are reformulated as (1) pre-contemplation, (2) pre-action [merged], and (3) action-maintenance. The threshold estimates are closely related to the proportion of responses in each stage (Anderson, 1984). However, the spacing of the threshold parameters also appears to reveal information about respondent preferences. In cases where threshold points are tightly bunched, we expect a lower level of discrimination to separate those stages behaviorally. Using biking as an example, on the one hand, the threshold locations indicate a higher likelihood of belonging in the early adoption stages, as also shown in Figure 1. Notwithstanding, the region for the merged *pre-preparation* cycling stage did not result from simply combining the smallest stage proportions. Rather, it reveals that the intermediate stage thresholds, despite being more narrow proportionally, are still behaviorally distinct based on the covariates used to model willingness to use active travel mode.

Given that the magnitude of threshold parameters cannot be directly interpreted we define contour maps to illustrate the implications of the thresholds under various scenarios. Results of this analysis are shown in Figure 3. The three vertical Panels in Fig. 3 represent 2-dimensional frontier plots derived by iteratively varying a pair of explanatory variables over roughly 100,000 runs for each plot, while the remainder of the model 5 features are kept constant. The color scheme shows stage transition patterns from early stages (red) to mature use (green) as a function of the set of explanatory variables. Notably, while each set of variables are not significant for each mode-decision, they do impact each stage membership indirectly through the captured correlation.

***ABOUT HERE* Figure 3**. 2-D Adoption frontier mapping for active travel decisions under three scenarios. *Rows: Mode adoption. Column A: Competing active mobility barriers. Column B: Competing active mobility assets; Column C: Multimodal identity*

In column panel A of Figure 3 we represent two conditions that are known to act as barriers for active mobility adoption (Cervero & Duncan 2003). The first column of contour plots illustrates the relative impacts of living in two-vehicle households with low schedule flexibility on each of the modes. The vertical patterns for biking and walking illustrates the importance of the household car-fleet, and the comparative lack of sensitivity to work schedule flexibility. For bikesharing the tendency to be in pre-contemplation, the most common experience in the sample, is unaffected by work-hour flexibility.

Column B represents the connections between two important features in promoting active travel, namely perceived pedestrian infrastructure quality and openness to learning. From the model results, we would expect walking to be impacted by the infrastructure and cycling to be connected to learning. However, the joint model generates more nuanced predictions. We note a stark horizontal threshold for walking, suggesting that infrastructure quality is indeed much more decisive than the mindset variable. Instead, for cycling, the learning mindset appears to drive the stage progression only up to a threshold. Notably, a positive view of both infrastructure quality openness to learning are needed to move from pre- contemplation to pre-action.

Column C shows the connection between multimodal experiences and declared active mobility identity, the latter of which is significant for each of the mode-adoption processes. The strong significance of the individual variables translates into well-defined adoption frontiers for each mode. Notably, bikesharing is driven mainly by variation in multimodality, and a high index score is needed to trigger a threshold transition. Walking is driven largely by a personal motivation to engage in active travel. Private cycling appears to be highly sensitive to both multimodality and self-identity to reach active user status.

An important take-away from the two-dimensional contour plots is the divergent impacts of the barriers, facilitators and multimodal/identity factors for each mode. It is important to point out that the contour mapping displays non-linearity that would be overlooked in separate stage-of-change modelling.

**6 Conclusion**

6.1. Summary and Discussion

With the rise of multimodal service options to match the convenience of private car ownership, there is a growing need to consider the adoption of different modes as parallel, rather than independent, processes. The objective of this paper is to investigate the relationships between active travel adoption behaviors, represented via stage-of-change model structures driven by a number of individual, psychological and neighborhood dynamics. Examining the multivariate ordered probit model results and the error term correlations, results point to the following general findings. First, walking and cycling represent complementary processes and thereby need to be viewed as a joint adoption process. The main linkage for these decisions is grounded in the active transportation identities and the sense of neighborhood belonging of travelers. The two cycling decisions (private cycling and bikesharing) are also viewed as a joint decision-making process, suggesting that regular bikesharing users will more readily consider owning and using their own bicycles in the future and vice versa. Finally, walking and bikesharing adoption processes are slightly substitutional, with weak negative correlation between the two processes. In conjunction with the significant predictor variables describing active travel mode adoption, these findings suggest guidelines for the theory-driven design of behavior change campaigns that encourage sustainability via a range of possible mobility service packages. Second, while the ordered stages of change definition in the joint model is well supported we find that the stages themselves are not uniform. Each active mode is shown to have a distinctive adoption path with either three or four separate stages and meanings. The implications of these differences, as well as the interdependence of active travel mobility modes, is illustrated by deriving two-dimensional contour plots for selected scenarios. Third, most important implication of the joint processes is the observed ripple effects of variables that run across active mode-adoption decisions. This is particularly noticeable for identity and norm variables that have joint significance. A novel finding to explore further is the practical implication of the stage-transitions. We provide initial evidence that threshold frontiers can be abrupt and highly responsive to joint features.

**6.2 Policy Implications**

This work has implications for understanding the emerging topic in the psychology literature of *behavioral spillover*. This phenomenon describes the situation in which the adoption of one behavior influences the likelihood of adopting additional behaviors that share a common goal, such as pro-environmentalism (Galizzi and Whitmarsh, 2019). For example, Evans et al. (2012) find that it is more effective to motivate car share users to recycle by targeting the shared goal of 'protecting the environment' as opposed to offering a monetary incentive. Lanzini and Thøgersen (2014), however, conclude not only that financial compensation strategies are more effective at encouraging both adoption and spillover compared to verbal reinforcement of 'green' values, but also that the initial target behavior induced other pro-environmental behaviors only when they were considered low-cost, low-effort actions. Thus, although critical for understanding the mechanisms of lifestyle change, the processes underpinning behavioral spillover are not well understood, and research has yet to illuminate sound policy recommendations, let alone consistent methodological guidelines (Galizzi and Whitmarsh, 2019).

Although analyzed at an aggregated level via the stages of change, this study produces evidence of potential pathways for *motivational spillover* effects within the same behavioral domain through a joint modeling framework. More specifically, commonalities among the ordered probit model components could indicate principal psychological mechanisms underlying the potential for 'positive spillover' in the domain of active mobility. That is, the adoption of one behavior would influence the adoption, or intention to adopt, a related behavior, as membership in a specific stage is a 'decision' whether or not a traveler is consciously aware of it. We conjecture that the multiple dimensions of identity explored in the survey provides the foundation for the linkage. To expound, *identity change* has been conjectured to be critical for promoting complementary behaviors in an effort for individuals to achieve consistency between their self-concept and past/future behaviors (Lauren et al., 2019; Nash et al., 2017), though there remains considerable ambiguity surrounding how to best design interventions around this construct (Carrico et al., 2018). Thus, by unifying stage-based analysis of active travel into a joint model framework, this paper gives new insight on the readiness for engaging in a new behavior, in addition to the corresponding sources of motivation to change.

### 6.3 Research Recommendations

Future research is needed to deepen and expand this work along the following dimensions. *First*, as done in (Becker et al., 2017), the probit formulation could be extended with the capacity to model any combination of dependent variable types, so that adoption behaviors could be explored jointly with, for instance, latent factor scores corresponding to constructs not included in the final models such as lifestyle factors. *Second*, the current models assume homogeneous threshold effects for the coefficients, which might be implausible given the notion of tailored policy that coincides with multi-stage behavior change theory. Future work should relax this assumption. Stage-specific effects could imply for example that multimodal experiences are decisive only for some stage transitions, not other adjacent stages. *Third*, data collection and modeling should account for a wider set of travel behaviors, along with the longitudinal and likely non-linear nature of the adoption process. In particular, it would be useful to demonstrate how the uptake of various sustainable travel modes is influenced by the ability and willingness to reduce the U.S. status quo of private car use, which has its own set of environmental and psychological determinants. Moreover, with panel data, it would be possible to capture true spillover at the individual level while tracking more in-depth stage transition patterns (i.e. instances of forward or backward membership change) across potentially interrelated behaviors. This would, as Dolan and Galizzi (2015) highlight, shed light on how the maintenance of multiple, possibly conflicting, identity goals might result in unexpected substitution patterns that should be considered when designing mobility behavior policies, e.g. *permitting* oneself to go for a leisure drive on the weekend (goal: achievement of social status) after cycling during the week for the work-home commute (goal: maintenance of physically active lifestyle). Finally, from the perspective of future data collection, the lack of random sampling coupled with the cross-sectional nature of the data hampers the capacity of the empirical findings to be generalizable.

**Acknowledgements**: Funding recognition to be competed for author(s). The study was approved by the University's Institutional Review Board. Any opinions, findings, and conclusions or recommendations expressed in this material are those of the authors and do not necessarily reflect the views of the National Science Foundation.

Conflicts of Interest: None

References

Ajzen, I., 1991. The theory of planned behavior. *Organizational behavior and human decision processes* 50(2), 179-211.

Anagnostopoulou, E., Bothos, E., Magoutas, B., Schrammel, J., Mentzas, G., 2018. Persuasive Technologies for Sustainable Mobility: State of the Art and Emerging Trends. *Sustainability* 10(7), 2128.

Anastasopoulos, P.C., Karlaftis, M.G., Haddock, J.E., Mannering, F.L., 2012. Household Automobile and Motorcycle Ownership Analyzed with Random Parameters Bivariate Ordered Probit Model. *Transportation Research Record*(2279), 12-20.

Anderson, J.A., 1984. Regression and Ordered Categorical Variables. *Journal of the Royal Statistical Society. Series B (Methodological)* 46(1), 1-30.

Andersson, A., Hiselius, L.W., Adell, E., 2018. Promoting sustainable travel behaviour through the use of smartphone applications: A review and development of a conceptual model. *Travel Behaviour and Society* 11, 52-61.

Armitage, C.J., 2009. Is there utility in the transtheoretical model? *British Journal of Health Psychology* 14(2), 195-210.

Astroza, S., Garikapati, V.M., Bhat, C.R., Pendyala, R.M., Lavieri, P.S., Dias, F.F., 2017. Analysis of the impact of technology use on multimodality and activity travel characteristics. *Transportation Research Record* 2666(1), 19-28.

Bamberg, S., 2013. Applying the stage model of self-regulated behavioral change in a car use reduction intervention. *Journal of Environmental Psychology* 33, 68-75.

Bamberg, S., Hunecke, M., Blöbaum, A., 2007. Social context, personal norms and the use of public transportation: Two field studies. *Journal of Environmental Psychology* 27(3), 190-203.

Becker, H., Loder, A., Schmid, B., Axhausen, K.W., 2017. Modeling car-sharing membership as a mobility tool: A multivariate Probit approach with latent variables. *Travel Behaviour and Society* 8, 26-36.

Bhat, C.R., 2015. A new generalized heterogeneous data model (GHDM) to jointly model mixed types of dependent variables. *Transportation Research Part B: Methodological* 79, 50-77.

Bhat, C.R., Guo, J.Y., 2007. A comprehensive analysis of built environment characteristics on household residential choice and auto ownership levels. *Transportation Research Part B: Methodological* 41(5), 506-526.

Bhat, C.R., Srinivasan, S., 2005. A multidimensional mixed ordered-response model for analyzing weekend activity participation. *Transportation Research Part B: Methodological* 39(3), 255-278.

Biehl, A., Stathopoulos, A., (2020) Investigating the interconnectedness of active transportation and public transit usage as a primer for Mobility-as-a-Service adoption and deployment, Journal of Transport & Health, Vol 18, https://doi.org/10.1016/j.jth.2020.100897

Biehl, A., Ermagun, A., Stathopoulos, A., 2018. Modelling determinants of walking and cycling adoption: A stage-of-change perspective. *Transportation Research Part F: Traffic Psychology and Behaviour* 58, 452-470.

Biehl, A., Ermagun, A., Stathopoulos, A., 2019. Utilizing multi-stage behavior change theory to model the process of bike share adoption. *Transport Policy* 77, 30-45.

Carfora, V., Caso, D., Sparks, P., Conner, M., 2017. Moderating effects of pro-environmental self-identity on pro-environmental intentions and behaviour: A multi-behaviour study. *Journal of Environmental Psychology* 53, 92-99.

Carrico, A.R., Raimi, K.T., Truelove, H.B., Eby, B., 2018. Putting Your Money Where Your Mouth Is: An Experimental Test of Pro-Environmental Spillover From Reducing Meat Consumption to Monetary Donations. *Environment and Behavior* 50(7), 723-748.

Cass, N., Faulconbridge, J., 2016. Commuting practices: New insights into modal shift from theories of social practice. *Transport Policy* 45, 1-14.

Castillo-Manzano, J.I., Castro-Nuño, M., López-Valpuesta, L., 2015. Analyzing the transition from a public bicycle system to bicycle ownership: A complex relationship. *Transportation Research Part D: Transport and Environment* 38, 15-26.

Cervero, R., Duncan, M., 2003. Walking, bicycling, and urban landscapes: evidence from the San Francisco Bay Area. *American Journal of Public Health* 93(9), 1478-1483.

Choo, S., Mokhtarian, P.L., 2008. How do people respond to congestion mitigation policies? A multivariate probit model of the individual consideration of three travel-related strategy bundles. *Transportation* 35(2), 145-163.

Chorus, C.G., 2015. Models of moral decision making: Literature review and research agenda for discrete choice analysis. *Journal of Choice Modelling* 16, 69-85.

Clark, A.F., Scott, D.M., 2013. Does the social environment influence active travel? An investigation of walking in Hamilton, Canada. *Journal of Transport Geography* 31, 278-285.

Copperman, R.B., Bhat, C.R., 2007. An analysis of the determinants of children's weekend physical activity participation. *Transportation* 34(1), 67.

Diana, M., Pirra, M., 2016. A comparative assessment of synthetic indices to measure multimodality behaviours. *Transportmetrica A: Transport Science* 12(9), 771-793.

Dias, F.F., Lavieri, P.S., Sharda, S., Khoeini, S., Bhat, C.R., Pendyala, R.M., Pinjari, A.R., Ramadurai, G., Srinivasan, K.K., 2020. A comparison of online and in-person activity engagement: The case of shopping and eating meals. *Transportation Research Part C: Emerging Technologies* 114, 643-656.

Dolan, P., Galizzi, M.M., 2015. Like ripples on a pond: Behavioral spillovers and their implications for research and policy. *Journal of Economic Psychology* 47, 1-16.

Evans, L., Maio, G.R., Corner, A., Hodgetts, C.J., Ahmed, S., Hahn, U., 2012. Self-interest and pro-environmental behaviour. *Nature Climate Change* 3, 122.

Ewing, R., Cervero, R., 2010. Travel and the Built Environment. *Journal of the American Planning Association* 76(3), 265-294.

Fielding, K.S., McDonald, R., Louis, W.R., 2008. Theory of planned behaviour, identity and intentions to engage in environmental activism. *Journal of Environmental Psychology* 28(4), 318-326.

Forward, S.E., 2014. Exploring people's willingness to bike using a combination of the theory of planned behavioural and the transtheoretical model. *Revue Européenne de Psychologie Appliquée/European Review of Applied Psychology* 64(3), 151-159.

French, S., Wood, L., Foster, S.A., Giles-Corti, B., Frank, L., Learnihan, V., 2014. Sense of Community and Its Association With the Neighborhood Built Environment. *Environment and Behavior* 46(6), 677-697.

Friman, M., Huck, J., Olsson, L., 2017. Transtheoretical Model of Change during Travel Behavior Interventions: An Integrative Review. *International Journal of Environmental Research and Public Health* 14(6), 581.

Galizzi, M.M., Whitmarsh, L., 2019. How to measure behavioral spillovers: a methodological review and checklist. *Frontiers in psychology* 10.

Garrow, L.A., Chen, Z., Ilbeigi, M., Lurkin, V., 2020. A new twist on the gig economy: conducting surveys on Amazon Mechanical Turk. *Transportation* 47(1), 23-42.

Gatersleben, B., Appleton, K.M., 2007. Contemplating cycling to work: Attitudes and perceptions in different stages of change. *Transportation Research Part a-Policy and Practice* 41(4), 302-312.

Golob, T.F., Regan, A.C., 2002. Trucking industry adoption of information technology: a multivariate discrete choice model. *Transportation Research Part C-Emerging Technologies* 10(3), 205-228.

Greene, W.H., Hensher, D.A., 2010. *Modeling ordered choices: A primer*. Cambridge University Press.

Guo, J.Y., Bhat, C.R., Copperman, R.B., 2007. Effect of the built environment an motorized and nonmotorized trip making - Substitutive, complementary, or synergistic? *Transportation Research Record*(2010), 1-11.

Hagger, M.S., Chatzisarantis, N.L.D., 2009. Integrating the theory of planned behaviour and self-determination theory in health behaviour: A meta-analysis. *British Journal of Health Psychology* 14(2), 275-302.

Handy, S., van Wee, B., Kroesen, M., 2014. Promoting Cycling for Transport: Research Needs and Challenges. *Transport Reviews* 34(1), 4-24.

Heineke, K., Kloss, B., Scurtu, D., Weig, F., 2019. Micromobility's 15,000-mile checkup, *McKinsey Insights*.

Heinen, E., 2018. Are multimodals more likely to change their travel behaviour? A cross-sectional analysis to explore the theoretical link between multimodality and the intention to change mode choice. *Transportation Research Part F: Traffic Psychology and Behaviour* 56, 200-214.

Heinen, E., Handy, S., 2012. Similarities in Attitudes and Norms and the Effect on Bicycle Commuting: Evidence from the Bicycle Cities Davis and Delft. *International Journal of Sustainable Transportation* 6(5), 257-281.

Heinen, E., Ogilvie, D., 2016. Variability in baseline travel behaviour as a predictor of changes in commuting by active travel, car and public transport: a natural experimental study. *Journal of Transport & Health* 3(1), 77-85.

Heinen, E., van Wee, B., Maat, K., 2010. Commuting by Bicycle: An Overview of the Literature. *Transport Reviews* 30(1), 59-96.

Hirk, R., Hornik, K., Vana, L., 2020. mvord: An R Package for Fitting Multivariate Ordinal Regression Models. *2020* 93(4), 41.

Johe, M.H., Bhullar, N., 2016. To buy or not to buy: The roles of self-identity, attitudes, perceived behavioral control and norms in organic consumerism. *Ecological Economics* 128, 99-105.

Kamargianni, M., Polydoropoulou, A., 2013. Hybrid choice model to investigate effects of teenagers' attitudes toward walking and cycling on mode choice behavior. *Transportation research record* 2382(1), 151-161.

Kees, J., Berry, C., Burton, S., Sheehan, K., 2017. An Analysis of Data Quality: Professional Panels, Student Subject Pools, and Amazon's Mechanical Turk. *Journal of Advertising* 46(1), 141-155.

Keizer, M., Sargisson, R.J., van Zomeren, M., Steg, L., 2019. When personal norms predict the acceptability of push and pull car-reduction policies: Testing the ABC model and low-cost hypothesis. *Transportation Research Part F: Traffic Psychology and Behaviour* 64, 413-423.

Keller, A., Eisen, C., Hanss, D., 2019. Lessons Learned From Applications of the Stage Model of Self-Regulated Behavioral Change: A Review. *Frontiers in Psychology* 10(1091).

Klöckner, C.A., Blöbaum, A., 2010. A comprehensive action determination model: Toward a broader understanding of ecological behaviour using the example of travel mode choice. *Journal of Environmental Psychology* 30(4), 574-586.

Knez, I., Eliasson, I., 2017. Relationships between Personal and Collective Place Identity and Well-Being in Mountain Communities. *Frontiers in Psychology* 8(79).

Kwak, N., Radler, B., 2002. A comparison between mail and web surveys: Response pattern, respondent profile, and data quality. *Journal of Official Statistics* 18(2), 257-273.

Langbroek, J.H.M., Franklin, J.P., Susilo, Y.O., 2016. The effect of policy incentives on electric vehicle adoption. *Energy Policy* 94, 94-103.

Lanzini, P., Thøgersen, J., 2014. Behavioural spillover in the environmental domain: an intervention study. *Journal of Environmental Psychology* 40, 381-390.

Lauren, N., Smith, L.D.G., Louis, W.R., Dean, A.J., 2019. Promoting Spillover: How Past Behaviors Increase Environmental Intentions by Cueing Self-Perceptions. *Environment and Behavior* 51(3), 235-258.

Lind, H.B., Nordfjærn, T., Jørgensen, S.H., Rundmo, T., 2015. The value-belief-norm theory, personal norms and sustainable travel mode choice in urban areas. *Journal of Environmental Psychology* 44, 119-125.

Lindhjem, H., Navrud, S., 2011. Are Internet surveys an alternative to face-to-face interviews in contingent valuation? *Ecological Economics* 70(9), 1628-1637.

Ma, L., Dill, J., Mohr, C., 2014. The objective versus the perceived environment: what matters for bicycling? *Transportation* 41(6), 1135-1152.

Mason, W., Suri, S., 2012. Conducting behavioral research on Amazon's Mechanical Turk. *Behavior Research Methods* 44(1), 1-23.

Moser, G., 2009. Quality of life and sustainability: Toward person–environment congruity. *Journal of Environmental Psychology* 29(3), 351-357.

Mulley, C., 2017. Mobility as a Services (MaaS) – does it have critical mass? *Transport Reviews* 37(3), 247-251.

Nash, N., Whitmarsh, L., Capstick, S., Hargreaves, T., Poortinga, W., Thomas, G., Sautkina, E., Xenias, D., 2017. Climate-relevant behavioral spillover and the potential contribution of social practice theory. *Wiley Interdisciplinary Reviews: Climate Change* 8(6), e481.

Nazari, F., Noruzoliaee, M., Mohammadian, A.K., 2018. Shared versus private mobility: Modeling public interest in autonomous vehicles accounting for latent attitudes. *Transportation Research Part C: Emerging Technologies* 97, 456-477.

Page, N.C., Nilsson, V.O., 2017. Active Commuting: Workplace Health Promotion for Improved Employee Well-Being and Organizational Behavior. *Frontiers in Psychology* 7(1994).

Parkes, S.D., Jopson, A., Marsden, G., 2016. Understanding travel behaviour change during mega-events: Lessons from the London 2012 Games. *Transportation Research Part A: Policy and Practice* 92, 104-119.

Pinjari, A.R., Eluru, N., Bhat, C.R., Pendyala, R.M., Spissu, E., 2008. Joint Model of Choice of Residential Neighborhood and Bicycle Ownership: Accounting for Self-Selection and Unobserved Heterogeneity. *Transportation Research Record* 2082(1), 17-26.

Pinjari, A.R., Pendyala, R.M., Bhat, C.R., Waddell, P.A., 2011. Modeling the choice continuum: an integrated model of residential location, auto ownership, bicycle ownership, and commute tour mode choice decisions. *Transportation* 38(6), 933.

Prochaska, J.O., Redding, C.A., Evers, K.E., 2008. The Transtheoretical Model and Stages of Change, In: Glanz, K., Rimer, B.K., Viswanath, K. (Eds.), *Health Behavior and Health Education: Theory, Research, and Practice.* Jossey-Bass, San Francisco, CA, pp. 97-121.

Prochaska, J.O., Velicer, W.F., 1997. The transtheoretical model of health behavior change. *American journal of health promotion* 12(1), 38-48.

Pucher, J., Dijkstra, L., 2000. Making walking and cycling safer: lessons from Europe. *Transportation Quarterly* 54(3), 25-50.

Pucher, J., Garrard, J., Greaves, S., 2011. Cycling down under: a comparative analysis of bicycling trends and policies in Sydney and Melbourne. *Journal of Transport Geography* 19(2), 332-345.

Rashidi, T.H., Mohammadian, A., Koppelman, F.S., 2011. Modeling interdependencies between vehicle transaction, residential relocation and job change. *Transportation* 38(6), 909.

Revelle, W., 2016. psych: Procedures for Psychological, Psychometric, and Personality Research. Northwestern University, Evanston, Illinios.

Riggs, W., 2017. Painting the fence: Social norms as economic incentives to non-automotive travel behavior. *Travel Behaviour and Society* 7, 26-33.

Ruiz, T., Bernabé, J.C., 2014. Measuring factors influencing valuation of nonmotorized improvement measures. *Transportation Research Part A: Policy and Practice* 67, 195-211.

Salon, D., 2015. Heterogeneity in the relationship between the built environment and driving: Focus on neighborhood type and travel purpose. *Research in Transportation Economics* 52, 34-45.

Scheiner, J., 2018. Why is there change in travel behaviour? In search of a theoretical framework for mobility biographies. *ERDKUNDE* 72(1), 41-62.

Schneider, R.J., 2013. Theory of routine mode choice decisions: An operational framework to increase sustainable transportation. *Transport Policy* 25, 128-137.

Scott, D.M., Kanaroglou, P.S., 2002. An activity-episode generation model that captures interactions between household heads: development and empirical analysis. *Transportation Research Part B: Methodological* 36(10), 875-896.

Sheehan, K.B., 2018. Crowdsourcing research: Data collection with Amazon's Mechanical Turk. *Communication Monographs* 85(1), 140-156.

Singleton, P.A., 2018. Walking (and cycling) to well-being: Modal and other determinants of subjective well-being during the commute. *Travel Behaviour and Society*.

Smith, N.A., Sabat, I.E., Martinez, L.R., Weaver, K., Xu, S., 2015. A Convenient Solution: Using MTurk To Sample From Hard-To-Reach Populations. *Industrial and Organizational Psychology* 8(2), 220-228.

StataCorp, 2017. Stata Statistical Software: Release 15. StataCorp LLC, College Station, TX.

Stern, P.C., Dietz, T., Abel, T., Guagnano, G.A., Kalof, L., 1999. A value-belief-norm theory of support for social movements: The case of environmentalism. *Human ecology review*, 81-97.

Tang, J., Zhen, F., Cao, J., Mokhtarian, P.L., 2018. How do passengers use travel time? A case study of Shanghai-Nanjing high speed rail. *Transportation* 45(2), 451-477.

Thigpen, C.G., Driller, B.K., Handy, S.L., 2015. Using a stages of change approach to explore opportunities for increasing bicycle commuting. *Transportation Research Part D: Transport and Environment* 39, 44-55.

Uzzell, D., Pol, E., Badenas, D., 2002. Place Identification, Social Cohesion, and Enviornmental Sustainability. *Environment and Behavior* 34(1), 26-53.

van der Werff, E., Steg, L., Keizer, K., 2013. The value of environmental self-identity: The relationship between biospheric values, environmental self-identity and environmental preferences, intentions and behaviour. *Journal of Environmental Psychology* 34, 55-63.

Walton, T.N., Jones, R.E., 2018. Ecological Identity: The Development and Assessment of a Measurement Scale. *Environment and Behavior* 50(6), 657-689.

Washington, S., Karlaftis, M.G., Mannering, F., Anastasopoulos, P., 2020. *Statistical and econometric methods for transportation data analysis.* CRC press.

Waygood, E.O.D., Avineri, E., 2016. Communicating transportation carbon dioxide emissions information: Does gender impact behavioral response? *Transportation Research Part D: Transport and Environment* 48, 187-202.

Whitmarsh, L., O'Neill, S., 2010. Green identity, green living? The role of pro-environmental self-identity in determining consistency across diverse pro-environmental behaviours. *Journal of Environmental Psychology* 30(3), 305-314.

Wilkerson, A., Carlson, N.E., Yen, I.H., Michael, Y.L., 2012. Neighborhood Physical Features and Relationships With Neighbors:Does Positive Physical Environment Increase Neighborliness? *Environment and Behavior* 44(5), 595-615.

Willis, D.P., Manaugh, K., El-Geneidy, A., 2015. Cycling Under Influence: Summarizing the Influence of Perceptions, Attitudes, Habits, and Social Environments on Cycling for Transportation. *International Journal of Sustainable Transportation* 9(8), 565-579.

Winters, M., Davidson, G., Kao, D., Teschke, K., 2011. Motivators and deterrents of bicycling: comparing influences on decisions to ride. *Transportation* 38(1), 153-168.

Yamamoto, T., Shankar, V.N., 2004. Bivariate ordered-response probit model of driver's and passenger's injury severities in collisions with fixed objects. *Accident Analysis and Prevention* 36(5), 869-876.

Zarif, R., Pankratz, D.M., Kelman, B., 2019. Small is beautiful: Making micromobility work for citizens, cities, and service providers, *Deloitte Insights.*

Zuniga, K.D., 2012. From barrier elimination to barrier negotiation: A qualitative study of parents' attitudes about active travel for elementary school trips. *Transport Policy* 20, 75-81.

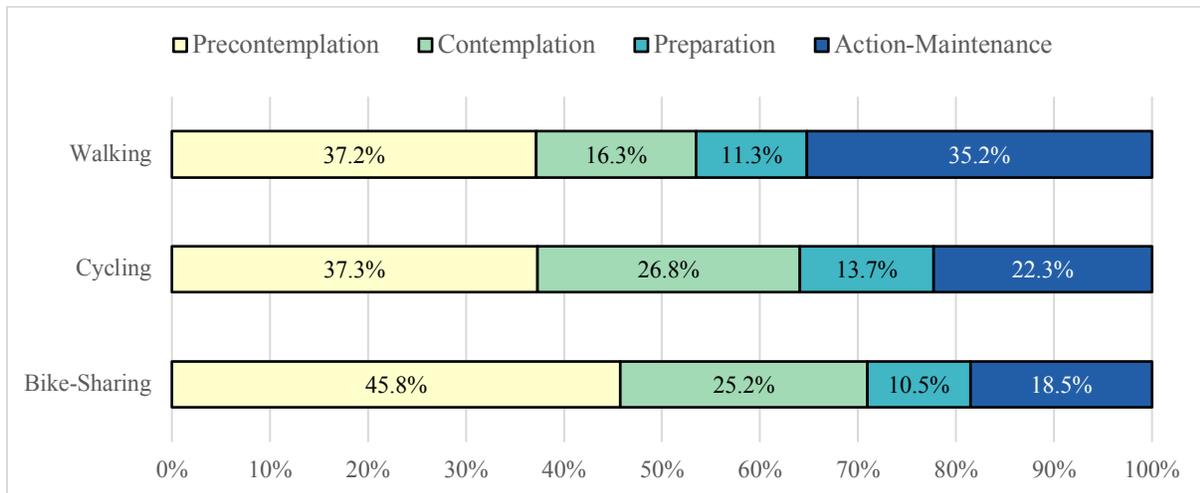

Figure 1. Stage-of-change membership breakdown for three active travel modes (n = 826).

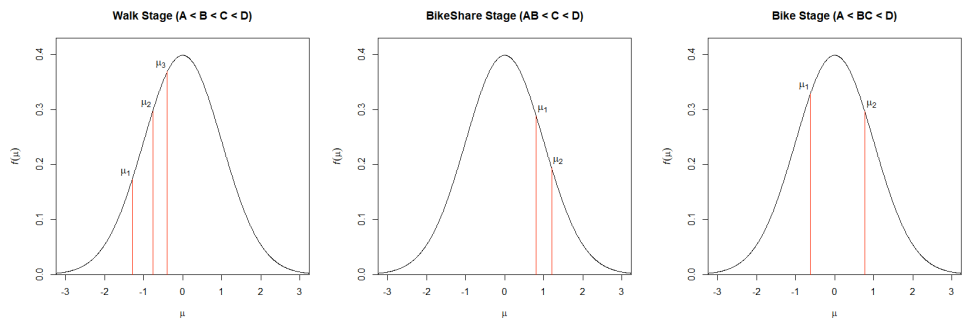

Figure 2. Final threshold structure for active mode adoption processes (Multivariate structure)

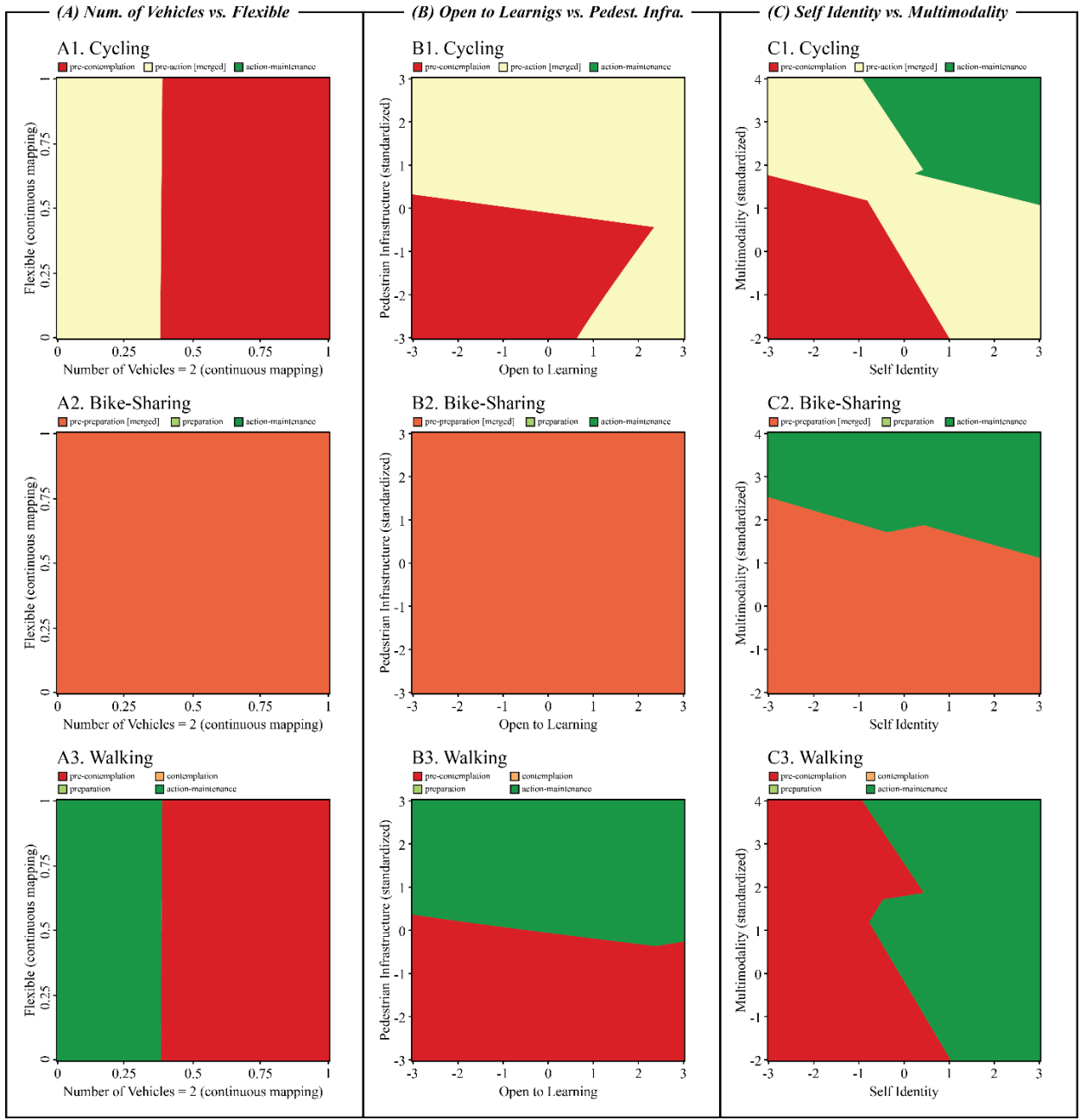

**ABOUT HERE** **Figure 3**. 2-D Adoption frontier mapping for active travel decisions under three scenarios. *Rows: Mode adoption. Column A: Competing active mobility barriers. Column B: Competing active mobility assets; Column C: Multimodal identity*

Table 1. Key characteristics of the survey sample (n = 826).

| Variables | Categories | # in Sample | % of Sample |
|---|---|---|---|
| *Socio-demographic* | | | |
| Age (4 NA) | 18-24 years | 139 | 16.8 |
| | 25-34 years | 294 | 35.6 |
| | 35-44 years | 187 | 22.6 |
| | 45-54 years | 100 | 12.1 |
| | 55+ years | 102 | 12.3 |
| Sex (4 NA) | Female | 483 | 58.5 |
| | Male | 339 | 41.0 |
| Race/Ethnicity (2 NA) | Non-Hispanic White | 684 | 82.8 |
| | Hispanic/Nonwhite/Mixed | 140 | 16.9 |
| Employment Status (52 NA) | Employed for wages | 591 | 71.5 |
| | Self-employed | 90 | 10.9 |
| | Homemaker | 39 | 4.7 |
| | Retired | 29 | 3.5 |
| | Unemployed | 25 | 3.0 |
| Student | Full-time | 62 | 7.5 |
| | Part-time | 32 | 3.9 |
| Education (4 NA) | No college degree | 244 | 29.5 |
| | Associate's or Bachelor's | 411 | 49.8 |
| | Graduate degree | 167 | 20.2 |
| Annual HH Income (22 NA) | $30K or less | 170 | 20.6 |
| | $30-50K | 187 | 22.6 |
| | $50-70K | 149 | 18.0 |
| | $70-90K | 120 | 14.5 |
| | Greater than $90K | 178 | 21.5 |
| HH Size (3 NA) | Live alone | 176 | 21.3 |
| | w/ another person | 281 | 34.0 |
| | w/ 3 or 4 people | 299 | 36.2 |
| | w/ more than 4 people | 67 | 8.1 |
| *Situational* | | | |
| State of Residence (56 NA) *Derived from zip code | Illinois | 272 | 32.9 |
| | Indiana | 63 | 7.6 |
| | Michigan | 145 | 17.6 |
| | Minnesota | 68 | 8.2 |
| | Ohio | 147 | 17.8 |
| | Wisconsin | 74 | 9.0 |
| Neighborhood Type | Urban | 217 | 26.3 |
| | Suburban | 321 | 38.9 |
| | Urb-Sub 'Hybrid' | 288 | 34.9 |
| *Household mobility* | | | |
| Driver's License | Yes | 772 | 93.5 |
| Transit Access | Yes | 705 | 85.4 |
| # HH Vehicles | 0 | 71 | 8.6 |
| | 1 | 299 | 36.2 |
| | 2 | 354 | 42.9 |
| | 3 or more | 102 | 12.3 |
| # HH Bicycles | 0 | 171 | 20.7 |
| | 1 | 257 | 31.1 |
| | 2 | 215 | 26.0 |
| | 3 or more | 183 | 22.2 |

Table 2. Names and definitions of probit model covariates besides the extracted factor variables.

| Nomenclature | Definition |
|---|---|
| Num Vehicle i | Number of vehicles owned by a household; if *i* or more, then the value is accompanied by a + sign<br>Base: 0 vehicles |
| Num Bicycle i | Number of bicycles owned by a household; if *i* or more, then the value is accompanied by a + sign<br>Base: 0 bicycles |
| No License | Dummy variable indicating that individual does not have a driver's license |
| Suburban | Dummy variable indicating that individual lives in a suburban area, in comparison to urban or 'hybrid' areas |
| Female | Dummy variable indicating that individual is female |
| Full-time Worker | Dummy variable indicating that individual is a full-time employee |
| Full-time Student | Dummy variable indicating that individual is full-time student |
| College Degree | Dummy variable indicating that individual has attained a college degree |
| Flexible | Dummy variable indicating that individual has some time flexibility in his or her schedule, compared to "maybe" or "no flexibility" |
| Travel Variety | "The idea of adding variety to my travel habits is appealing to me."<br>1 = Strongly disagree to 5 = Strongly agree |
| Travel Satisfaction | "I am satisfied with the choices I make regarding my travel."<br>1 = Strongly disagree to 5 = Strongly agree |
| Travel Boredom | "I tend to feel bored while traveling."<br>1 = Strongly agree to 5 = Strongly disagree |
| Street Infrastructure | Perceived quality of local street infrastructure<br>1 = Terrible to 5 = Excellent |
| Pedest. Infrastructure | Perceived quality of local pedestrian infrastructure<br>1 = Terrible to 5 = Excellent |
| Mental Map | "I am confident in my knowledge of where places are in my neighborhood and how to get to them."<br>1 = Strongly disagree to 7 = Strongly agree |
| Multimodal | Adjusted Shannon entropy index for multimodality<br>Continuous: 0=unimodal, …, 1=multimodal |

Table 3. *Independently Estimated Ordered probit models for biking, walking and bike-share*

| *Model Statistics* | Cycling | | Bikesharing | | Walking | |
|---|---|---|---|---|---|---|
| *Number of Observations* | *826* | | *826* | | *826* | |
| *Log-Likelihood at zero* | *-1096.69* | | *-1035.95* | | *-1054.94* | |
| *Final Log-Likelihood* | *-932.19* | | *-948.93* | | *-904.08* | |
| *$\rho^2$* | *0.150* | | *0.084* | | *0.143* | |
| *AIC* | *1,896.38* | | *1,915.86* | | *1,848.16* | |
| *BIC* | *1,971.85* | | *1,958.31* | | *1,942.49* | |
| **Parameter** | **Coefficient** | **z-value** | **Coefficient** | **z-value** | **Coefficient** | **z-value** |
| Num Vehicle 1 | -0.657 | -4.26 | | | -0.813 | -4.36 |
| Num Vehicle 2 | -0.993 | -6.17 | | | -1.037 | -5.50 |
| Num Vehicle 3+ | -0.891 | -4.72 | | | -0.919 | -4.38 |
| Num Bicycle 1 | 0.901 | 7.49 | | | | |
| Num Bicycle 2 | 1.002 | 7.91 | | | | |
| Num Bicycle 3+ | 1.066 | 7.59 | | | | |
| No License | | | | | 0.886 | 4.00 |
| Suburban | -0.318 | -3.48 | | | -0.287 | -3.17 |
| Female | -0.263 | -3.18 | | | | |
| Full-Time Worker | | | | | -0.223 | -2.35 |
| Full-Time Student | | | | | 0.546 | 3.01 |
| College Degree | | | | | 0.233 | 2.44 |
| Flexible | | | 0.200 | 2.08 | | |
| Travel Variety | 0.167 | 3.60 | 0.134 | 2.93 | | |
| Travel Boredom | | | | | 0.116 | 2.60 |
| Street Infrastructure | | | | | -0.120 | -2.56 |
| Pedestrian Infrastructure | | | | | 0.212 | 4.37 |
| Mental Map | | | | | 0.120 | 2.76 |
| Self Identity | 0.129 | 3.06 | 0.129 | 3.01 | 0.131 | 2.98 |
| Place Identity | 0.090 | 2.05 | 0.160 | 3.66 | | |
| Social Identity | | | | | 0.123 | 2.90 |
| Personal Norms | | | 0.197 | 4.41 | 0.220 | 4.83 |
| Life Satisfaction | | | | | -0.128 | -2.89 |
| Open to Learning | 0.126 | 2.81 | | | | |
| Multimodal | 0.337 | 7.14 | 0.325 | 7.38 | 0.194 | 4.14 |
| **Thresholds** | Coefficient | z-value | Coefficient | z-value | Coefficient | z-value |
| $\mu_{(PC,C)}$ | -0.685 | -4.43 | 0.036 | 0.42 | -1.300 | -6.54 |
| $\mu_{(C,P)}$ | 0.194 | 1.32 | 0.792 | 8.87 | -0.759 | -3.87 |
| $\mu_{(P,AM)}$ | 0.719 | 4.97 | 1.185 | 12.65 | -0.376 | -1.91 |

Table 4. Trivariate probit model for stages-of-change for all three modes.

| Model Statistics: | Number of Observations: 826 | | $\rho^2$: 0.149 | | | AIC: 9,004.21 | | |
|---|---|---|---|---|---|---|---|---|
| | Log-Likelihood at zero: -5,185.21 | | Horowitz' $\rho^2$ vs. single modes: significant | | | BIC: 9,425.70 | | |
| | Final Log-Likelihood: -4,412.74 | | | | | | | |

| Parameter | Cycling | | | Bikesharing | | | Walking | | |
|---|---|---|---|---|---|---|---|---|---|
| | Coefficient | z-value | p-value | Coefficient | z-value | p-value | Coefficient | z-value | p-value |
| Num Vehicle 1 | -0.613 | -3.66 | 0.000 | | | | -0.778 | -3.95 | 0.000 |
| Num Vehicle 2 | -0.973 | -5.62 | 0.000 | | | | -1.021 | -5.10 | 0.000 |
| Num Vehicle 3+ | -0.818 | -3.93 | 0.000 | | | | -0.878 | -3.85 | 0.000 |
| Num Bicycle 1 | 0.923 | 6.91 | 0.000 | | | | | | |
| Num Bicycle 2 | 1.034 | 7.17 | 0.000 | | | | | | |
| Num Bicycle 3+ | 1.099 | 7.02 | 0.000 | | | | | | |
| No License | | | | | | | 0.945 | 4.51 | 0.000 |
| Suburban | -0.313 | -3.51 | 0.000 | | | | -0.378 | -4.17 | 0.000 |
| Female | -0.254 | -3.07 | 0.002 | | | | | | |
| Full-Time Worker | | | | | | | -0.216 | -2.29 | 0.022 |
| Full-Time Student | | | | | | | 0.496 | 2.89 | 0.004 |
| College Degree | | | | | | | 0.234 | 2.42 | 0.016 |
| Flexible | | | | 0.232 | 1.91 | 0.056 | | | |
| Travel Variety | 0.169 | 3.84 | 0.000 | 0.115 | 2.00 | 0.045 | | | |
| Travel Boredom | | | | | | | 0.089 | 2.12 | 0.034 |
| Street Infrastructure | | | | | | | -0.109 | -2.44 | 0.015 |
| Pedestrian Infrastructure | | | | | | | 0.219 | 4.43 | 0.000 |
| Mental Map | | | | | | | 0.095 | 2.19 | 0.029 |
| Self Identity | 0.125 | 2.85 | 0.004 | 0.106 | 2.19 | 0.029 | 0.154 | 3.57 | 0.000 |
| Place Identity | 0.091 | 2.15 | 0.032 | 0.108 | 2.20 | 0.028 | | | |
| Social Identity | | | | | | | 0.142 | 3.24 | 0.001 |
| Personal Norms | | | | 0.130 | 2.42 | 0.016 | 0.220 | 4.91 | 0.000 |
| Life Satisfaction | | | | | | | -0.098 | -2.14 | 0.033 |
| Open to Learning | 0.095 | 2.16 | 0.031 | | | | | | |
| Multimodal | 0.301 | 6.95 | 0.000 | 0.402 | 7.61 | 0.000 | | | |
| **Thresholds** | Coefficient | z-value | p-value | Coefficient | z-value | p-value | Coefficient | z-value | p-value |
| $\mu_{(PC,C)}$ | -0.620 | -3.54 | 0.000 | 0.809 | 7.33 | 0.000 | -1.295 | -6.07 | 0.000 |
| $\mu_{(C,P)}$ | 0.772 | 4.14 | 0.000 | | | | -0.768 | -3.65 | 0.000 |
| $\mu_{(P,AM)}$ | | | | 1.202 | 10.57 | 0.000 | -0.395 | -1.89 | 0.059 |
| **Correlations** | Coefficient | z-value | p-value | | | | | | |
| $\rho_{cycling,bikesharing}$ | 0.181 | 2.83 | 0.005 | | | | | | |
| $\rho_{cycling,walking}$ | 0.439 | 10.79 | 0.000 | | | | | | |
| $\rho_{bikesharing,walking}$ | -0.097 | -2.24 | 0.025 | | | | | | |

**Appendix**

*Table A1. Delineation of the stages of change for three active travel modes.*

*Table A2. Three-factor solution for Active Travel Disposition scale (loadings above 0.45 are bolded).*

*Table A3. Two-factor solution for Environmental Spatial Ability scale (loadings above 0.45 are bolded).*

*Table A4. Four-factor solution for Sense of Community scale (loadings above 0.45 are bolded).*

*Table A5. Four-factor solution for Psychological Well-Being scale (loadings above 0.45 are bolded).*

Table A1. Delineation of the stages of change for three active travel modes.

| Identification Steps | Response | Result | Stage |
|---|---|---|---|
| ***Walking and Cycling*** | | | |
| Which statement best describes your average weekly walking/cycling behavior as a primary mode of travel, considering all travel purposes? | I have never contemplated making a routine trip using this mode | See Row 2 | --- |
| | I have contemplated making a routine trip using this mode | See Row 3 | --- |
| | I use this mode for at least one routine trip | See Row 4 | --- |
| Is walking/cycling as the primary means of travel a realistic alternative for any routine trip? | No | Assignment | Precontemplation 1 |
| | Yes | Assignment | Precontemplation 2 |
| Do you expect to use walking/cycling as the primary means of travel for a routine trip in the near future? | No | Assignment | Contemplation |
| | Yes | Assignment | Preparation |
| How long have you been walking/cycling for a routine trip? | Less than one year | Assignment | Action |
| | One year or longer | Assignment | Maintenance |
| ***Bikesharing*** | | | |
| Assuming "good weather," would you expect to use bike share at least once per week? | Yes | Assignment | Action-Maintenance |
| | No | Next Step | --- |
| Would you ever contemplate using this mode? | Yes | Next Step | --- |
| | No | Assignment | Precontemplation |
| Is bike share currently accessible to you?<br><br>~ & ~<br><br>What is the likelihood of using bike share in the next six months? (5-point Likert scale) | No & 1-2 on Likert scale | Assignment | Contemplation 1 |
| | Yes & 1-2 on Likert scale | Assignment | Contemplation 2 |
| | No & 3-5 on Likert scale | Assignment | Preparation 1 |
| | Yes & 3-5 on Likert scale | Assignment | Preparation 2 |

Table A2. Three-factor solution for Active Travel Disposition scale (loadings above 0.45 are bolded).

| Statement: Strongly disagree to Strongly agree | Personal Norms | Place Identity | Self Identity |
|---|---|---|---|
| 1. It weighs on my conscience if I do not use active transportation for a trip when it is a reasonable alternative. | **0.630** | 0.171 | 0.244 |
| 2. Finding more opportunities to travel using active transportation is meaningful to me. | **0.831** | 0.282 | 0.232 |
| 3. I think it is right to take advantage of opportunities for me to travel using active transportation. | **0.664** | 0.333 | 0.263 |
| 4. I feel I should attempt to integrate more trips by active transportation into my weekly travel patterns. | **0.679** | 0.269 | 0.327 |
| 5. Investment in active transportation infrastructure in my neighborhood would make me feel valued in society. | 0.407 | **0.662** | 0.361 |
| 6. Investment in active transportation infrastructure would distinguish my neighborhood from others nearby. | 0.214 | **0.684** | 0.296 |
| 7. Increased availability of active transportation would make me more capable of traveling where I need. | 0.360 | **0.514** | 0.348 |
| 8. Increased availability of active transportation would create a neighborhood that aligns more with how I view myself. | 0.395 | **0.576** | **0.503** |
| 9. Greater popularity of active transportation could make me feel pressured to change how I travel. | 0.228 | 0.306 | **0.459** |
| 10. Greater popularity of active transportation would give me a greater sense of pride in my neighborhood. | 0.307 | **0.556** | **0.603** |
| 11. Participating in more active transportation would allow me to adhere more strongly to personal values. | 0.423 | 0.378 | **0.720** |
| 12. Participating in more active transportation would increase my confidence in being able to enjoy my ideal lifestyle. | 0.440 | 0.430 | **0.651** |
| **SS Loadings** | **3.004** | **2.525** | **2.398** |
| **Proportion of variance explained** | **0.250** | **0.210** | **0.200** |
| **Cumulative variance explained** | **0.250** | **0.461** | **0.661** |
| **Cronbach's alpha = 0.94** | | | |
| **Tucker Lewis index = 0.971** | | | |
| **RMSEA index = 0.063** | | | |

Table A3. Two-factor solution for Environmental Spatial Ability scale (loadings above 0.45 are bolded).

| Statement: Strongly agree to Strongly disagree | Spatial Orientation | GPS Tech Affinity |
|---|---|---|
| 1. I am good at giving directions.* | **0.805** | -0.102 |
| 2. I easily get lost when traveling in an unfamiliar area. | **0.810** | -0.166 |
| 3. I have trouble understanding directions. | **0.825** | -0.072 |
| 4. I am good at reading maps.* | **0.764** | -0.097 |
| 5. I prefer someone else to do the travel planning for trips in unfamiliar areas. | **0.692** | -0.068 |
| 6. I do not have a good mental map of my local environment. | **0.573** | -0.031 |
| 7. I could easily travel to a new location without on-the-go access to GPS technology.* | **0.726** | -0.298 |
| 8. I am confident in my abilities to use GPS technology.* | 0.328 | **0.568** |
| 9. It is important to be able to access information on the Internet while traveling. | -0.134 | **0.626** |
| 10. I enjoy trying out new routes to familiar destinations.* | **0.491** | 0.150 |
| 11. I am easily stressed when I feel lost during travel. | **0.609** | -0.130 |
| 12. More often than not, I depend on GPS technology to help me travel. | -0.374 | **0.712** |
| 13. GPS technology has allowed for more variety in my everyday travel. | -0.075 | **0.747** |
| 14. I feel that I get more accomplished because of technology. | -0.103 | **0.738** |
| **SS Loadings** | **4.795** | **2.510** |
| **Proportion of variance explained** | **0.342** | **0.179** |
| **Cumulative variance explained** | **0.342** | **0.522** |
| **Cronbach's alpha = 0.87** | | |
| **Tucker Lewis index = 0.948** | | |
| **RMSEA index = 0.062** | | |

*Indicates reverse scoring

Table A4. Four-factor solution for Sense of Community scale (loadings above 0.45 are bolded).

| Statement: Strongly agree to Strongly disagree | Community Cohesion | Descriptive Norms | Social Identity | Confidence |
|---|---|---|---|---|
| 1. Neighborhood members and I value the same things. | **0.644** | 0.279 | 0.195 | 0.256 |
| 2. Being a member of my neighborhood makes me feel good. | **0.689** | 0.158 | 0.386 | 0.201 |
| 3. I put a lot of time and effort in being a part of my neighborhood. | 0.312 | 0.229 | **0.766** | 0.187 |
| 4. Being a member of my neighborhood is an important part of my identity. | 0.307 | 0.246 | **0.767** | 0.210 |
| 5. I fit in very well with the people in my neighborhood. | **0.672** | 0.230 | 0.335 | 0.232 |
| 6. I enjoy interacting with other neighborhood residents. | **0.558** | 0.178 | 0.440 | 0.237 |
| 7. Local development trends make me feel more confident about the future of my neighborhood. | 0.335 | 0.274 | 0.208 | **0.704** |
| 8. Local development trends make me feel more confident about my own future. | 0.267 | 0.213 | 0.234 | **0.863** |
| 9. My neighborhood has symbols of membership such as signs, art, architecture, logos, and landmarks that people recognize. | 0.280 | **0.472** | 0.129 | 0.115 |
| 10. People in my neighborhood have similar needs, priorities, and goals. | **0.567** | 0.347 | 0.159 | 0.198 |
| 11. People in my neighborhood embrace innovation in transportation services. | 0.225 | **0.758** | 0.150 | 0.149 |
| 12. The local government successfully meets the needs of my neighborhood. | 0.388 | **0.481** | 0.068 | 0.228 |
| 13. Growth in active transportation use is a priority in my neighborhood. | 0.084 | **0.778** | 0.251 | 0.162 |
| **SS Loadings** | **2.636** | **2.176** | **1.898** | **1.688** |
| **Proportion of variance explained** | **0.203** | **0.167** | **0.146** | **0.130** |
| **Cumulative variance explained** | **0.203** | **0.370** | **0.516** | **0.646** |
| **Cronbach's alpha = 0.92** | | | | |
| **Tucker Lewis index = 0.963** | | | | |
| **RMSEA index = 0.060** | | | | |

Table A5. Four-factor solution for Psychological Well-Being scale (loadings above 0.45 are bolded).

| Statement: Strongly disagree to Strongly agree | Life Satisfaction | Open to Learning | Perseverance | Autonomy |
|---|---|---|---|---|
| 1. In most ways my life is close to ideal. | **0.894** | 0.089 | 0.135 | 0.063 |
| 2. I am satisfied with my life. | **0.917** | 0.086 | 0.132 | 0.020 |
| 3. So far I have been able to obtain the things I want in life. | **0.858** | 0.124 | 0.120 | 0.036 |
| 4. If I could live my life over, I would change almost nothing. | **0.636** | 0.079 | 0.070 | 0.164 |
| 5. My decisions are usually not influenced by what everyone else is doing. | 0.108 | 0.064 | 0.183 | **0.630** |
| 6. I judge myself by what I think is important, not by the values of what others think is important. | 0.103 | 0.154 | 0.137 | **0.612** |
| 7. In general, I feel I am in charge of what is happening in my life. | **0.578** | 0.225 | 0.157 | 0.282 |
| 8. I have been able to build a healthy lifestyle that is much to my liking. | **0.626** | 0.243 | 0.224 | 0.127 |
| 9. I am interested in activities that could give me a new perspective in life. | 0.134 | **0.780** | 0.059 | 0.090 |
| 10. I enjoy being in new situations that require me to rethink my habits. | 0.130 | **0.726** | 0.113 | 0.049 |
| 11. My life has been a continuous process of learning, changing, and growth. | 0.218 | 0.422 | 0.299 | 0.249 |
| 12. I do not set ambitious goals for myself because I am afraid of failure.* | 0.267 | 0.177 | **0.648** | 0.176 |
| 13. When trying to learn something new, I tend to give up if I am not initially successful.* | 0.138 | 0.134 | **0.780** | 0.213 |
| 14. I try to learn new things, even when they look too difficult for me. | 0.106 | 0.438 | 0.346 | 0.256 |
| **SS Loadings** | **3.715** | **1.729** | **1.435** | **1.113** |
| **Proportion of variance explained** | **0.265** | **0.123** | **0.103** | **0.080** |
| **Cumulative variance explained** | **0.265** | **0.389** | **0.491** | **0.571** |
| **Cronbach's alpha = 0.87** | | | | |
| **Tucker Lewis index = 0.970** | | | | |
| **RMSEA index = 0.047** | | | | |

*Indicates reverse scoring